\documentclass[sigconf,nonacm]{acmart}
\AtBeginDocument{%
  }

\setcopyright{none}
\copyrightyear{}
\acmYear{}
\acmDOI{}
\acmConference[Conference acronym 'XX]{Make sure to enter the correct
  conference title from your rights confirmation email}{X}{X}
\acmISBN{}

\usepackage{adjustbox}
\usepackage{xcolor}
\usepackage{multirow} 
\usepackage{enumitem}
\usepackage{booktabs}
\usepackage{graphicx}
\usepackage{subcaption}
\usepackage[most]{tcolorbox}
\usepackage[table]{xcolor} 
\usepackage{tabularx} 
\usepackage{placeins} 
\usepackage{xspace}   
\usepackage{caption}
\usepackage{makecell}
\usepackage{array}
\definecolor{lightgray}{gray}{0.9} 
\begin{document}
%
%
%
\newcommand{\chatbot}{WaLLM}
\newcommand{\chatbotactiveusers}{96}
\newcommand{\chatbottotalmsgs}{17,408}
\newcommand{\chatbottotalsessions}{3,952}
\newcommand{\pointsfrequent}{frequent leaderboard}
\newcommand{\pointsfrequentcaps}{Frequent leaderboard}
\newcommand{\pointsrare}{occasional leaderboard}
\newcommand{\pointsrarecaps}{Occasional leaderboard}
\newcommand{\servicefrequent}{regular}
\newcommand{\servicefrequentcaps}{Regular}
\newcommand{\servicerare}{casual}
\newcommand{\servicerarecaps}{Casual}
\newcommand{\ui}[1]{\textsf{#1}\xspace}

\newcommand{\term}[1]{\textbf{#1}\xspace}
\newcommand{\emphasize}[1]{\textbf{#1}\xspace}

\newcommand{\quoted}[1]{`#1'\xspace}

\newenvironment{userquote}
  {\begin{quotation}\small\noindent}
  {\end{quotation}}
%
%
\title[\chatbot: Use and Engagement with LLMs on WhatsApp]{\chatbot{}: Understanding Use and Engagement with a General-Purpose LLM on WhatsApp}
\author{Hiba Eltigani}
\affiliation{
  \institution{Tufts University}
  \country{USA}
  \city{Medford}
}

\author{Rukhshan Haroon}
\affiliation{
  \institution{Tufts University}
  \country{USA}
  \city{Medford}
}

\author{Asli Kocak}
\affiliation{
  \institution{Tufts University}
  \country{USA}
  \city{Medford}
}

\author{Abdullah Bin Faisal}
\affiliation{
  \institution{Tufts University}
  \country{USA}
  \city{Medford}
}

\author{Noah Martin}
\affiliation{
  \institution{Tufts University}
  \country{USA}
  \city{Medford}
}

\author{Fahad Dogar}
\affiliation{
  \institution{Tufts University}
  \country{USA}
  \city{Medford}
}
\renewcommand{\shortauthors}{Eltigani et al.}

%
\begin{abstract}
Large language model (LLM) chatbots are increasingly reaching users through messaging platforms (e.g. WhatsApp). However, these systems remain largely proprietary and opaque, while academic research has focused on narrow, domain-specific assistants. This leaves open questions about how people use general-purpose LLMs and how such systems should be designed. To address this gap, we developed~\chatbot{}, a general-purpose LLM chatbot, and deployed it on WhatsApp as a design probe to study open-ended AI use in the wild. Our findings show that health and well-being accounted for the largest proportion of queries, suggesting that users turned to~\chatbot{} for advice and information. Engagement features varied in their adoption and associated patterns of use: proactive communication supported the service's visibility and correlated with higher user activity, while communal lists facilitated content discovery. We report how these features were adapted to WhatsApp's affordances and discuss implications for designing general-purpose LLM services over messaging platforms.
\end{abstract}
%



\settopmatter{printacmref=false}
\setcopyright{none}
\renewcommand\footnotetextcopyrightpermission[1]{}
\pagestyle{plain}
\maketitle
\section{Introduction} 
Large Language Model (LLM)-based chatbots have broadened access to AI for non-technical users~\cite{johnnyprompt}. Their integration into messaging platforms highlights these platforms' growing role as entry points for AI use~\cite{indiaemergentusers,vashistha2019examining,whatsapp2b}. WhatsApp, with over three billion active users~\cite{whatsapp2b}, has been a common platform in prior HCI research for delivering AI-enabled services~\cite{ashabot,wheelpedia}. These deployments, while valuable, are typically in narrow, domain-specific tasks and do not fully capture the breadth of open-ended, everyday AI use.

At the other end of the spectrum, general-purpose assistants such as Meta AI~\cite{meta_ai}, OpenAI's ChatGPT~\cite{open_ai_whatsapp_bot}, and Perplexity~\cite{perplexityapi2024} have been offered over WhatsApp. These commercial systems are challenging to study because their interaction logs, design decisions, and underlying engagement strategies are largely proprietary. Existing work, therefore, leaves a gap between task-specific research deployments and general-purpose commercial assistants whose use and design remain comparatively opaque. Notably, WhatsApp moved to restrict third-party general-purpose chatbots in January 2026, leading OpenAI and Perplexity to retire their WhatsApp assistants~\cite{openairetiring,techcrunchretiring,euthreatns}.

Beyond this gap, WhatsApp's mobile-first, message-oriented interface, limited persistent interface elements, and integration into users' everyday conversations can shape how users interact with services delivered through it~\cite{whatfutures}. These dynamics are consequential for general-purpose LLMs, where users do not arrive with a single predefined task. Interaction conventions common to standalone AI applications, such as dedicated discovery interfaces, are not natively supported within WhatsApp. This calls for careful design that adapts both core interaction features and engagement mechanisms to the constraints of the messaging environment.

\begin{figure}
  \centering
  \includegraphics[width=0.35\textwidth]{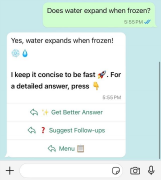}
  \caption{A WhatsApp user asking a question and \chatbot{} responding, with buttons that support usability (e.g., \ui{Menu}) and engagement (e.g., \ui{Suggest Follow-ups}).}
  \label{fig:example-interaction}
  \Description{Example of a WhatsApp user asking a question and \chatbot{} responding with an LLM-generated answer. The response message has additional buttons that facilitate usability (e.g., ``Menu'') and engagement (e.g., ``Suggest Follow-ups'').} 
\end{figure}
To study these dynamics, we developed and deployed~\chatbot{}, a general-purpose LLM chatbot on WhatsApp, as a design probe for studying open-ended use in the wild. Our deployment predates WhatsApp's 2026 restriction on third-party general-purpose chatbots, capturing use during a window that has since closed. Users interact with~\chatbot{} by sending a WhatsApp message, just as they would to a regular contact (Figure~\ref{fig:example-interaction}). With~\chatbot{}, we also experiment with adapting four engagement mechanisms to WhatsApp's constraints: communal question lists, proactive push messages, a contribution leaderboard, and follow-up suggestions. Our implementation examines the trade-offs introduced by these constraints rather than advocating for a single design approach.

Two research questions guide our empirical analysis:
\begin{enumerate}[leftmargin=*]
    \item \textbf{Patterns of Use:} What do users seek from a general-purpose LLM delivered through WhatsApp, and how do they interact with it?
    \item \textbf{Engagement Mechanisms:} How can established engagement mechanisms be adapted to fit WhatsApp's affordances, and how do users engage with these adaptations across different usage and interaction behaviors?
\end{enumerate}

In answering these questions, we find that health and well-being represent the largest proportion of queries, suggesting users turned to~\chatbot{} for advice and information. Most inaccurate responses went unchallenged, indicating that users may not always recognize or question incorrect information, a concern given how often they brought high-stakes topics such as health to the service. Engagement features revealed distinct patterns of adoption: proactive communication supported visibility and correlated with higher activity, while communal lists mainly supported content discovery among less engaged users. Follow-up suggestions were rarely used, but when users selected one, they overwhelmingly chose the two shown by default, underscoring the role of visibility in WhatsApp's constrained interface. To summarize, we make the following key contributions:
\begin{enumerate}[leftmargin=*]
    \item An in-the-wild account of open-ended LLM use over WhatsApp that characterizes users' queries and interactions.
    \item Empirical evidence on the transfer of engagement mechanisms to WhatsApp, including how users engage across different patterns of use.
    \item Design lessons for delivering general-purpose LLM services through WhatsApp.
\end{enumerate}
\section{Related Work}\label{sec:related_work}
Our work lies at the intersection of generative AI and information access. It draws on prior research on WhatsApp as a platform for broadening access to AI-enabled services and on the use of general-purpose LLMs. It also builds on work on engagement mechanisms for sustaining user participation in online systems.

\subsection{LLM-based services on WhatsApp}
WhatsApp is well suited to resource-constrained contexts: it supports text-based, multilingual communication, consumes minimal bandwidth, and runs reliably on low-end devices~\cite{indiaemergentusers}. Because of its ubiquity~\cite{whatsapp2b}, it has become a common platform for HCI researchers deploying socio-technical interventions~\cite{prasad-a-hybrid,varanasi2021tag,yadav-should-i,poon-engaging-high,karusala-that-courage}. These strengths informed our choice to deploy~\chatbot{} through WhatsApp, enabling users to interact with an LLM through tools already part of their daily lives.

The LLM-based systems examined in academic research remain largely domain-specific. For example, researchers have built a WhatsApp chatbot supporting community health workers in India \cite{ashabot}, and another providing wheelchair information in Kenya and Nigeria \cite{wheelpedia}. These systems demonstrate the promise of LLM-based services on WhatsApp but are tailored to narrow information needs.

In contrast to these research deployments, major AI companies have launched general-purpose LLM-based chatbots on WhatsApp~\cite{chatgpt_wtsp_usage_limit,meta_ai}. Meta AI, particularly, benefits from platform-native privileges (e.g., streaming responses) that enable richer features and deeper integration than those afforded to independent or research-based deployments. Most existing insights into LLM use come from the providers of proprietary platforms~\cite{peopleusegpt,anthropicclio,microsoftusage}. Across these studies, LLM interactions are examined through web browsers or dedicated apps. 

In this study, we investigate in-the-wild interactions with an LLM delivered over WhatsApp: a setting that surfaces leisure-oriented, everyday uses differing from the work-centric patterns of prior work.

\subsection{Engagement mechanisms in online platforms}
Prior research has explored strategies to increase participation in online systems; techniques such as points, leaderboards, badges, reminders, and suggested questions have been used with varying degrees of success~\cite{cheng2020building,whatfutures,replysuggestions,ke2026invisible,ashabot}. For example, a shared-response leaderboard ranking participants has supported a gamified youth-engagement exercise over WhatsApp~\cite{whatfutures}. Proactive messages such as reminders are often delivered through channels already embedded in users' routines, such as SMS or messaging apps. A scoping review of digital physical activity interventions found that such delivery improved reach and access, but participation tended to peak after scheduled contacts and decline in between, with overall engagement falling over time~\cite{ke2026invisible}.

Engagement also depends on users knowing what to ask. Non-expert users often struggle to formulate effective queries during chatbot interactions~\cite{nguyen-user,wheelpedia,johnnyprompt}. Suggested questions offer one way to reduce this burden. In web search, suggestions that anticipate a user's next step are preferred over merely relevant ones~\cite{rosset2020leading}, and in a WhatsApp-based chatbot for community health workers, related-question suggestions accounted for the majority of users' questions~\cite{ashabot}. Uptake of suggestions nonetheless varies widely with context, and suggestion buttons can increase the likelihood that users follow a chatbot's advice~\cite{replysuggestions}.

The rise of LLMs also appears to be reshaping which forms of participation endure: transactional knowledge communities such as Stack Overflow have seen declines in traffic and contribution, while more socially oriented spaces such as Reddit have remained vibrant~\cite{aiconsequencescommunities}. This suggests that social and communal engagement mechanisms may be particularly relevant for AI-based services~\cite{wang2026bridging}. Transferring such mechanisms to a new context is not straightforward, however: it requires both design choices that implement the desired mechanism and evidence of its effectiveness in the new setting. We study open-ended AI use within WhatsApp, focusing on the design requirements for adapting multiple engagement strategies to this setting.
\section{Overview of \chatbot} \label{sec:design}
\begin{figure*}[!t]
\centering
\begin{subfigure}[b]{0.24\textwidth}
  \centering
  \includegraphics[width=\linewidth]{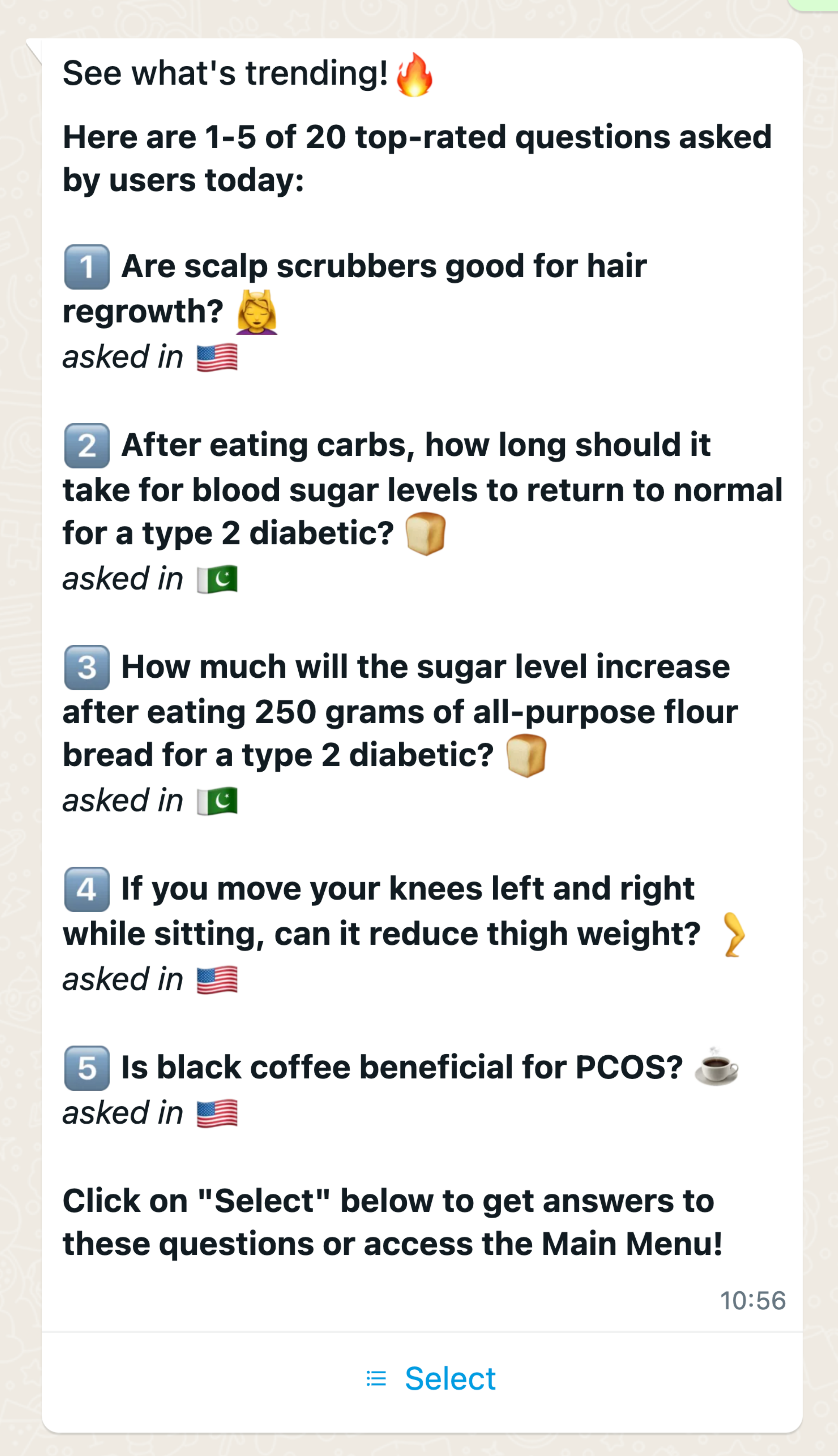} 
  \caption{\ui{Top-rated Questions}}
  \Description{Example of \ui{Top-rated Questions} message}
  \label{fig:navigation_trending}
\end{subfigure}
\hfill
\begin{subfigure}[b]{0.24\textwidth}
  \centering
  \includegraphics[width=\linewidth]{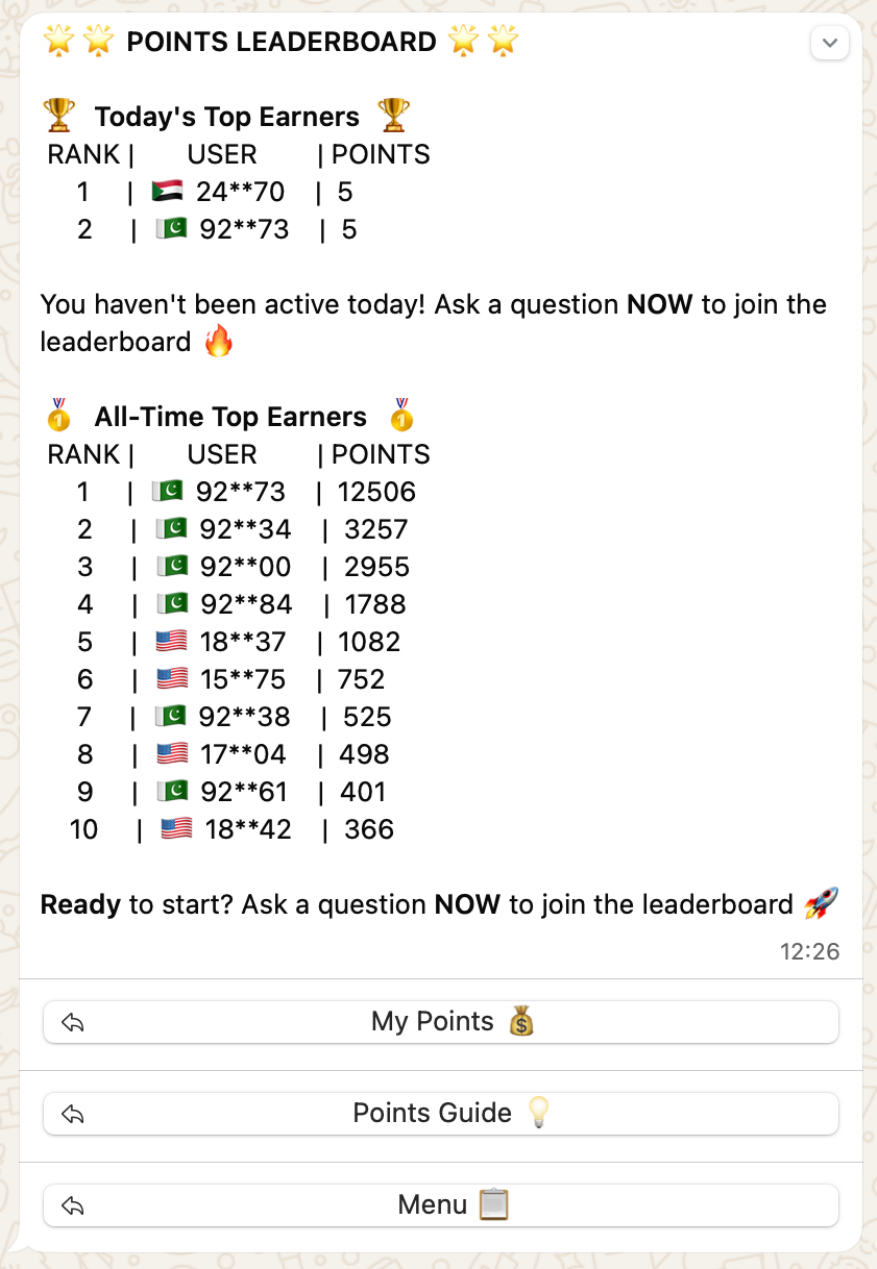}
  \caption{The \ui{Leaderboard}}
  \Description{Example of the \ui{Leaderboard} message}
  \label{fig:feature_leaderboard}
\end{subfigure}
\hfill
\begin{subfigure}[b]{0.24\textwidth}
  \centering
  \includegraphics[width=\linewidth]{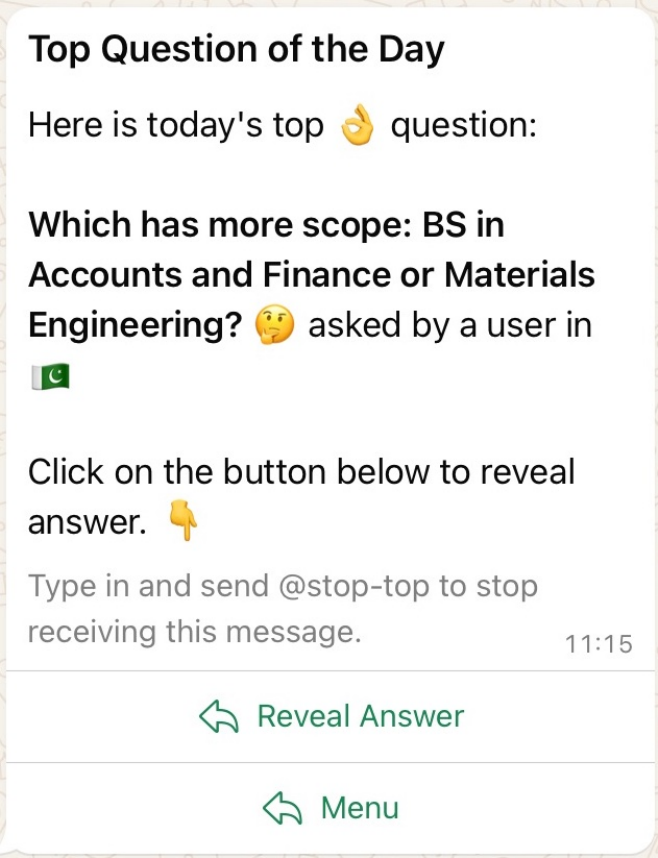} 
  \caption{\ui{Top Question of the Day}}
  \Description{Example of \ui{Top Question of the Day} message}
  \label{fig:navigation_topq_fig}
\end{subfigure}
\hfill
\begin{subfigure}[b]{0.24\textwidth}
  \centering
  \includegraphics[width=\linewidth]{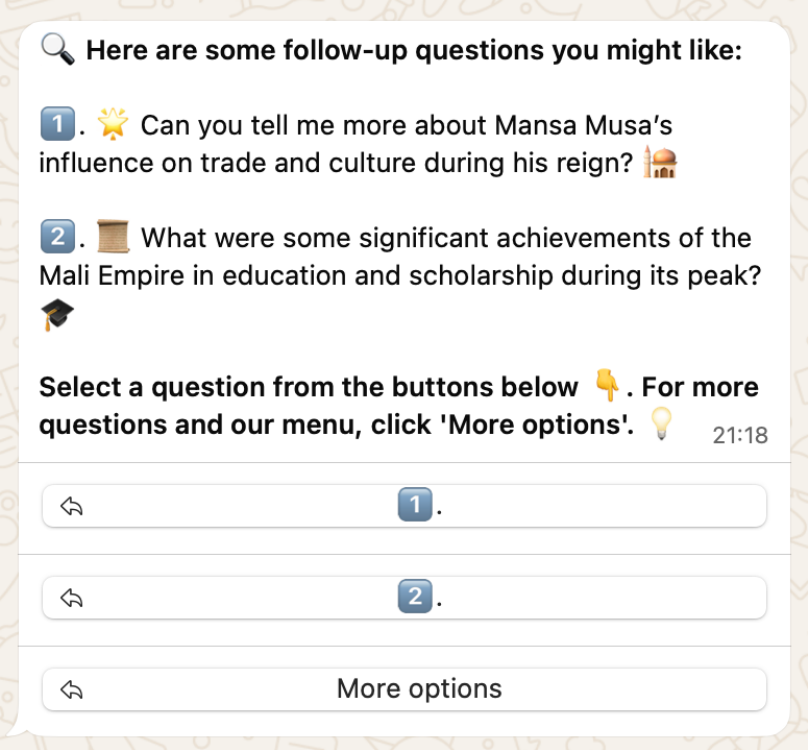} 
  \caption{\ui{Suggest Follow-ups}}
  \Description{Example of \ui{Suggest Follow-ups} message}
  \label{fig:navigation_suggest_initial}
\end{subfigure}
\caption{Examples of \chatbot's interactive messages}
\label{fig:feature_rewards_service}
\end{figure*}
\subsection{Design} \label{design} \label{navigation}
~\chatbot{} intentionally supports only text messages, the modality that LLMs natively handle and fully support through APIs. Text is also low-overhead, making it an affordable entry point for users with limited resources~\cite{indiaemergentusers,vashistha2019examining}. Users interact with~\chatbot{} either by typing queries or selecting from interactive messages, which offer AI-generated follow-up suggestions and curated questions previously asked by others. For clarity, we refer to typed queries as~\term{freeform} queries and selections via interactive messages as~\term{interactive} queries. 

WhatsApp's limited interface affordances shaped our core design. We implemented a button-based navigation interface to approximate persistent UI elements. It also allows users to move through long responses split across messages, mimicking response streaming on other platforms. In addition, the interface supports feature discovery and access, serving as a backbone for AI-generated query recommendations and engagement strategies.

Every response from~\chatbot{} includes three buttons in addition to the message text, as shown in Figure~\ref{fig:example-interaction}. The bottom-most button is persistently assigned to the \ui{Menu} item providing access to the service's features. The middle button, \ui{Suggest Follow-ups}, is for accessing AI-generated suggestions. The top-most button changes based on the situation; it is either used to request the next part in a long response (\ui{Continue Reading} button) or an alternative answer from a more powerful LLM (\ui{Get Better Answer} button).

\subsubsection{Communal Lists}\label{querylists} Many users' questions in a service like \chatbot{} are broadly relevant to other users, creating opportunities for communal reuse.~\chatbot{} curates two lists: \ui{Top-rated Questions} and \ui{Recent Questions} from users' questions. When a user submits a freeform query, an LLM evaluates whether it is (1) a question, (2) in English, and (3) self-contained. Eligible queries are added to the \ui{Recent Questions} list, and those with broader relevance, such as common experiences, general guidance, hobbies, or current events, are also included in the \ui{Top-rated Questions} list. The LLM additionally makes minor edits to align queries with WhatsApp's conversational norms (e.g., correcting errors, adding emojis~\cite{emojiuse}). Shared queries are anonymized to preserve user privacy (Figure~\ref{fig:navigation_trending}). Both lists are accessible through the \ui{Menu} button.

\subsubsection{Contribution Leaderboard}\label{rewardssystem}~\chatbot{} incorporates a reward system to give users points for contributing to the community by asking questions and interacting with the service. Users can use two views to monitor their standing. \ui{My Points} view summarizes total points, ranking position, and point breakdown by query category, and the \ui{Leaderboard} view displays the top 10 users over the past 24 hours and since launch (Figure~\ref{fig:feature_leaderboard}).

\subsubsection{Proactive Communication}\label{TopQdesign}Because WhatsApp prioritizes recent conversations, inactive conversations can quickly lose visibility, motivating proactive communication to resurface the service.~\chatbot{} uses WhatsApp template messages to send push messages. Besides resurfacing conversations, these messages can be used to announce updates or collect feedback.~\chatbot{} implemented the \ui{Top Question of the Day} (TopQ): a daily message featuring a question drawn from the \ui{Top-rated Questions} list, with quick access to both its answer and the main menu (Fig.~\ref{fig:navigation_topq_fig}). TopQ brings \chatbot's conversation back to the top of users' chat lists while sharing interesting questions from other users. All push messages included an opt-out option to preserve user autonomy.

\subsubsection{AI-generated Follow-up Suggestions}\label{suggestions} 
To reduce the difficulty of formulating queries,~\chatbot{} automatically generates related follow-up questions for each query (\ui{Suggest Follow-ups}). Six suggested questions are generated per query with responses to these suggestions pre-fetched to minimize latency and enhance user experience. When suggestions are requested,~\chatbot{} replaces the standard button layout with a suggestions view in which the top two buttons correspond to the first two of the six questions for quick access, and the third reveals the remaining questions (Figure~\ref{fig:navigation_suggest_initial}).

\subsection{Implementation} 
\chatbot{} is hosted on Amazon Web Services (AWS)~\cite{aws}, integrated with WhatsApp Business API~\cite{whatsapp}, and built on a research orchestration system~\cite{llmbridge2025}. The Business API supports the custom interactive messages (e.g., buttons, lists) described in Section~\ref{design}. The orchestration system is cloud-based and leverages a combination of OpenAI GPT models (e.g. GPT-4~\cite{openai2024gpt4}) and Anthropic Claude models~\cite{anthropic} to support core functionalities such as generating answers, suggesting follow-up questions, and adding WhatsApp-style elements. Both the underlying model and the prompts (see Appendix~\ref{appendix-prompts} for details) used to generate responses are hidden from users. The primary model used for response generation is regularly updated to account for new releases and cost reductions. Additional models are used selectively to support tasks such as query curation and alternative response generation.
\section{Methods}\label{sec:methods}
\subsection{Deployment and User Population}
\chatbot{} was deployed over a 15-month period. Users signed up for~\chatbot{} by sending a designated introductory message over WhatsApp and agreeing to the service's terms and conditions. Access to the system was open to adult, English-speaking users. The terms and conditions were delivered via WhatsApp in English. Users were not given any onboarding or usage instructions.

Users were recruited from Pakistan and Sudan, as well as African diaspora communities in the United States via snowball sampling in two waves; one early during the launch and another during a workshop event 11 months after that. The deployment reached these communities for practical access reasons and due to their familiarity with WhatsApp. We do not treat these participants as representative and we make no regional claims. In this paper, we focus on the launch cohort (96 users), cutting off the data at the second recruitment event (month 11) to avoid contaminating it with the second cohort (32 users), who were introduced under different conditions.

\subsection{Data Collection}
Our dataset consists of all user interactions with ~\chatbot. Each event is timestamped and assigned a unique identifier that indicates the feature or interface element used. The dataset also includes events captured via the WhatsApp Business API, such as message statuses (e.g., delivered, read), for users who have enabled this setting on their devices.

\subsection{Ethical Considerations}
All study procedures were conducted in accordance with human subjects research standards and were approved by our Institutional Review Board (IRB). Users were required to be at least 18 years old and to accept the service's terms and privacy policy before using~\chatbot. To protect user privacy, we applied restricted-access controls, content filtering, and data anonymization. Integration with WhatsApp was supported through end-to-end message encryption.

\subsection{Data Analysis} \label{dataAnalysis}
We adopted a mixed-methods approach combining quantitative analysis of service logs with qualitative analysis of user queries.

\subsubsection{Quantitative Analysis} 
Quantitative analysis was conducted on service logs using descriptive statistics and statistical tests. User interactions were grouped into sessions, defined as sequences of consecutive events with no gap exceeding 15 minutes~\cite{bardAnalysis}. We analyzed patterns of feature use and user engagement across sessions, focusing on features with high usage or design relevance.

\subsubsection{Qualitative Analysis} \label{qualitative}
Qualitative analysis focused on a subset of user queries collected during the first three months after deployment, a period that accounted for 63.8\% of all launch-cohort queries and included every user in the cohort. From this period we drew a stratified sample based on users' total number of sessions, randomly selecting 10\% of queries. The resulting sample covered 57\% of the cohort's users and preserved the underlying participation imbalance, in which 80\% of queries were submitted by only 6\% of users.

We employed thematic coding~\cite{braun2006using}, using an initial coding framework informed by prior work on query intent~\cite{wang2024understanding,cai2020predicting} and existing taxonomies~\cite{cambazoglu2021intent}, and extended it with additional codes derived from the data where necessary. Queries were coded along four dimensions: (a) query purpose, (b) type of information required for an answer, (c) whether the answer could be objectively verified, and (d) user intent. Using the same sample, we separately evaluated~\chatbot's corresponding responses as acceptable or unacceptable, following the criteria in Section~\ref{freeform-accuracy}.

A member of the research team imported the queries and responses to NVivo, applying thematic codes to the queries and accuracy ratings to the responses. The query codes were then grouped into categories, which were discussed with other members of the team. A second member who is not part of the initial study team independently validated the categories and associated codes and queries, along with response ratings. Any disagreements were resolved through discussion until consensus was reached. A similar approach was adopted by other researchers~\cite{haroon-twips-a,ahsen2022designing}.
\section{Findings}\label{sec:findings}
\begin{table*}[t]
  \centering
  \caption{Categories of users based on activity}
  \label{tab:users_cat_activity}
\begin{tabular}{@{}lccc@{}}
\toprule
\multicolumn{1}{l}{}& \multicolumn{1}{l}{One-time users} & \multicolumn{1}{l}{\servicerarecaps{} users} & \multicolumn{1}{l}{\servicefrequentcaps{} users} \\ \midrule
\# of sessions per user & 1   & 2 - 100 & more than 100  \\
\# of users & 17 & 71 & 8  \\
Total sessions & 17 & 706 & 3,229 \\
Total interactions  & 54  & 2,820   & 14,534 \\
Total freeform queries & 22  & 927  & 4,722 \\
Total interactive queries  & 29  & 1,720  & 8,117 \\
Average active days & 1  & 8  & 133\\
Average idle time between sessions (hours) & NA   & 172  & 16 \\
Average session duration in minutes  & 2.5   & 2.5 & 3.7  \\ \bottomrule
\end{tabular}
\end{table*}
\chatbotactiveusers{} users engaged with~\chatbot{} at least once through freeform or interactive queries, resulting in~\chatbottotalmsgs{} interactions across~\chatbottotalsessions{} sessions. Based on session count, we identify three user groups (Table~\ref{tab:users_cat_activity}): one-time users (18\%),~\servicerare{} users (74\%) and~\servicefrequent{} users (8\%). Thresholds were set heuristically from the observed distribution. ~\servicefrequentcaps{} users were substantially more engaged (133 active days on average) and accounted for 83\% and 82\% of total freeform and interactive queries respectively. For the remainder of this section, we examine the nature of user queries in Section~\ref{findings-freeform}; and analyze feature use and engagement in Section~\ref{findings-engagement-features}. 

\begin{table}[htbp]
  \centering
  \caption{Distribution of User Queries by Category (Sampled Data)}
  \label{tab:subjects_cat}
\begin{tabular}{@{}cc@{}}
  \toprule
  \makecell[c]{Category}  & \makecell[c]{\% of Queries}\\
  \midrule
  Health and Well-being                      & 28\%\\
  Cultural and General Knowledge & 17\%\\ 
  Science and Technology                     & 11\%\\
  Language and Communication                 & 10\%\\
  Social and Personal Development            & 9\%\\
  Commerce and Economy                       & 8\%\\   
  Education and Learning                     & 7\%\\
  Animals and Nature                         & 3\%\\
  Chatbot Information                        & 2\%\\
  Uncategorized                              & 5\%\\
  \bottomrule
\end{tabular}
\end{table}
\subsection{Characterizing User Queries and Interaction Patterns}\label{findings-freeform}
\subsubsection{Query Topics and Intent}
Users asked about a wide range of subjects (Table~\ref{tab:subjects_cat}). The two largest categories were health and well-being, and cultural and general knowledge. Queries related to managing diseases, eating right, exercise, and self-care accounted for 28\% of freeform queries (e.g. \quoted{Why my gums start aching... occasionally}\footnote{User quotes are reproduced verbatim, including original spelling and grammar.}). Users asked ~\chatbot{} to create personalized meal plans, verify medical information, and provide general health guidance. Cultural and general knowledge (e.g. \quoted{What are the 7 wonders of the world?}) constituted 17\% of freeform queries, but was concentrated during the early period of the deployment before winding down. Health and well-being remained prominent throughout the deployment.

Over 90\% of queries were information-seeking questions; while the remainder were statements serving other purposes, such as correcting \chatbot{} or clarifying intent (Table~\ref{tab:freeform_categories}). Of these questions, 58\% required factual knowledge, 31\% were non-factual and sought a general perspective, and 11\% were ambiguous due to insufficient context or spelling errors. Personal advice was a major part of the non-factual category: 50\% of non-factual questions were consultations on topics, such as health and family relations.

\servicerarecaps{} and \servicefrequent{} users showed no substantial difference in the questions they selected from communal lists; for both, the most-chosen items were non-factual, Social and Personal Development questions such as \quoted{How can we provide emotional care for our parents in their old age?} (see Appendix~\ref{appendix-trending-questions}).

Factual and advice-seeking questions dominated from users' first session. During the first session, 76\% of users who sent freeform queries asked factual questions, and 18\% sought personal advice.

\begin{table*}[t]
\centering
\caption{Categories of freeform queries\textsuperscript{*}}
\label{tab:freeform_categories}

\small
\setlength{\tabcolsep}{3pt}
\renewcommand{\arraystretch}{1.12}

\begin{tabularx}{\textwidth}{
  @{}
  >{\raggedright\arraybackslash}p{1.35cm}
  >{\raggedright\arraybackslash}p{1.65cm}
  >{\raggedright\arraybackslash}p{2.15cm}
  >{\raggedright\arraybackslash}X
  >{\centering\arraybackslash}p{1.65cm}
  >{\centering\arraybackslash}p{1.55cm}
  @{}
}
\toprule
\parbox[t]{1.35cm}{Purpose} &
\parbox[t]{1.65cm}{Type} &
\parbox[t]{2.15cm}{Intent} &
Example &
\parbox[t]{1.65cm}{\centering \% in First\\Session (n)} &
\parbox[t]{1.55cm}{\centering \% of Sample\\(n)} \\
\midrule

Question & Factual & Information
& What is excise duty?
& 44\% (79)
& 31\% (112) \\

& & Verification
& can they put MRNA vaccine in vegetables?
& 8\% (15)
& 12\% (43) \\

& & Explanation
& how to source high quality bedding sheets for bedding company business startup?
& 6\% (11)
& 12\% (44) \\

\addlinespace[0.25em]

& Non-factual & Advice
& how long after gall stone surgery I can travel?
& 11\% (19)
& 14\% (52) \\

& & Viewpoint
& what are the cons of being the oldest child?
& 18\% (32)
& 9\% (34) \\

& & \makecell[l]{Communication\\and Language}
& fancy words for exercise?
& 1\% (2)
& 5\% (19) \\

\addlinespace[0.25em]

& Ambiguous & Ambiguous
& zodiac sign
& 4\% (7)
& 11\% (38) \\

\midrule

Statement & & &
I meant the book `in the line of fire'..
& 8\% (14)
& 5\% (18) \\

\midrule

\multicolumn{4}{c}{Total}
& 100\% (179)
& 99\% (360) \\

\bottomrule

\multicolumn{6}{@{}p{\textwidth}@{}}{\footnotesize
\textsuperscript{*} Percentages may not sum to 100\% due to rounding.}

\end{tabularx}
\end{table*}
\subsubsection{Query Characteristics} \label{findings-characteristics}
Most freeform queries were in English, with only 2\% in other languages such as Arabic or Urdu. Those users first asked \chatbot{} whether it could understand the foreign language before switching to it. The majority of queries were well-formed sentences; the remaining 11\% consisted of keyword lists. Separately, 9\% of freeform queries contained spelling errors.

A further characteristic concerned how self-contained queries were. About 22\% of freeform queries required additional context; 9\% depended on information shared earlier in the session (e.g. \quoted{can you give me a range}), and 13\% had multiple interpretations whose resolution depended on earlier exchanges (e.g. \quoted{what would be a good multivitamin in my case?}). Full conversation history was rarely needed: the immediately preceding query-response pair was sufficient in 63\% of these cases, no relevant prior conversation existed in a further 36\%, and only 1\% required two prior pairs. 

\subsubsection{Accuracy of \chatbot's Responses}\label{freeform-accuracy}
LLMs are known to hallucinate, generating plausible but false or unsupported content~\cite{ji2023survey}. We evaluated each response in the sampled dataset as acceptable or unacceptable (see Section~\ref{qualitative} for method). A response was acceptable if it was accurate and addressed the user's intent (a refusal counted as acceptable when it explained why~\cite{gricemaxims}), and unacceptable if it was inaccurate or presented outdated information in vague terms that prevented the user from recognizing the limitation.

About 9\% of responses were unacceptable, concerning current news, product details, online content, local context, and \chatbot's own design. To understand users' responses to these cases, we treated any explicit signal of dissatisfaction such as correcting \chatbot, rephrasing the query, or response regeneration as an indication of user doubt. Nearly two-thirds of unacceptable responses were not challenged. One possible explanation is that users were not closely scrutinising answers while initially testing the system, in which case such instances would be expected to occur early in a user's interaction history. However, most uncontested cases occurred well into users' histories, suggesting that initial experimentation accounts for only a small portion of them.

\begin{table}
  \centering
  \caption{Distribution of interactive queries per user group across interactive features}
  \label{tab:interactive_usage_groups}
\begin{tabular}{@{}llccc@{}}
\toprule
Feature                   & One Time & \servicerarecaps & \servicefrequentcaps \\
\midrule
Continue Reading (Freeform Queries)          & 7\%    & 23\%           & 25\%  \\
Top-rated Questions       & 62\%   & 25\%           & 14\%  \\
Recent Questions          & 0\%    & 3\%            & 13\%  \\
Contribution Leaderboard  & 4\%    & 8\%            & 16\%  \\
Suggest Follow-ups        & 10\%   & 16\%           & 9\%   \\
Get Better Answer         & 17\%   & 25\%           & 23\%  \\
\bottomrule
\end{tabular}
\end{table}
\subsection{Feature Use and Engagement}\label{findings-engagement-features}
\subsubsection{Interactive Query Patterns Across User Groups} 
We observed distinct differences across user groups in their engagement with \chatbot's features. Table~\ref{tab:interactive_usage_groups} shows how use of features is distributed within each group, instead of absolute values which differ substantially across groups. This use includes activities related to a specific feature, such as navigating within a feature and reading responses to its related messages. In addition, TopQ messages activity is accounted for within \ui{Top-rated Questions} feature. 

One-time users' activity was dominated by the curated \ui{Top-rated Questions} list (62\% of interactions); they never used \ui{Recent Questions}. Although \ui{Get Better Answer} was their next-largest activity, they viewed none of the returned responses.~\servicefrequentcaps{} users showed the opposite emphasis; \ui{Top-rated Questions} fell to 14\%, while \ui{Recent Questions} rose to 13\%. Their activity centered on freeform queries: reading responses to their own queries accounted for the largest share (25\%), followed by generating alternative answers to them (23\%). ~\servicerarecaps{} users, on the other hand, split their attention almost evenly between \ui{Top-rated Questions} (25\%), \ui{Get Better Answer} (25\%), and freeform queries reading (23\%). Overall, newcomers relied on curated lists, while established users navigated recent and self-generated content.

\subsubsection{Proactive Communication}\label{topQfindings}
TopQ messages were sent daily to all users for nearly two months, beginning three weeks after the service launch (62 messages in total). On average, days on which a TopQ message was sent had twice as many active users than on days without one. However, because message days were not randomized, this association does not by itself establish causation.

To further examine this pattern, we analyzed 47 users, excluding new users due to their initially higher activity and one-time users due to their lack of multi-session history. On average, 63\% of users' active days occurred within 24 hours of a TopQ message, and 33 of the 47 had their final session during the broadcast period.

We also examined changes in activity patterns in one-hour intervals during the 24 hours following each TopQ message. Activity spiked within the first hour, and retrieving an answer to a TopQ message was the most common way users initiated a session within that hour. Across the full 24-hour window, 44\% of sessions began with a freeform query, whereas 56\% began with an interactive query. This differs noticeably from other periods, when 84\% of sessions began with freeform queries.  

Our analysis cannot distinguish between users who read push messages without responding and those who did not read them; thus, their actual impact may be underreported due to muted notifications or archived conversations. 

\subsubsection{Contribution Leaderboard} The reward feature was accessed by 38\% of users, with adoption varying across user groups: 75\% of \servicefrequent{} users, 41\% of \servicerare{} users, and  6\% of one-time users used the feature at least once. Interactions were heavily concentrated on the \ui{Leaderboard} rather than \ui{My Points} (10:1). 

Because \ui{Leaderboard} access dominated, we focused our analysis on users who accessed it. Among these users, 43\% viewed it only once, while 34\% accessed it multiple times within a single session. We therefore focused on the users with repeated access, classified as \term{\pointsrare} users (9 users) and \term{\pointsfrequent} users (5 users). To be classified as a \pointsfrequent, a user had to meet both criteria: (1) repeated access within a session and (2) access in more than 10 per cent of their total sessions.

Given the small size of these groups, the following comparisons are descriptive. For this analysis, interaction counts exclude reward-service (\ui{Leaderboard} and \ui{My Points}) and \ui{Menu} interactions, so that counts reflect use of the service rather than navigation. \pointsfrequentcaps{} users (n=5) had a median of 829 interactions, compared with 52 for \pointsrare{} users (n=9). Consistent with this higher activity, four of the five \pointsfrequent{} users (80\%) appeared on the \emphasize{all-time} top list, compared with two of the nine \pointsrare{} users (22\%). Across these 14 users, sessions in which the \ui{Leaderboard} was accessed had higher interaction counts (Wilcoxon Signed-Rank Test: \textit{W} = 4.5, \textit{p} $=$ 0.002, n $=$ 14 paired users). Because users chose when to access the \ui{Leaderboard}, we cannot distinguish whether its use encouraged further engagement or simply co-occurred with already engaged sessions.

\pointsfrequentcaps{} users adopted the feature within their first two sessions and continued using it consistently. \pointsrarecaps{} users were exclusively \servicerare{} users, while \pointsfrequent{} users included both \servicerare{} and \servicefrequent{} users.

\subsubsection{Follow-up Suggestions}\label{suggested_finding}
\chatbot's \ui{Suggest Follow-ups} feature accounted for 6.6\% of interactions and 11.8\% of interactive queries. Of the \chatbotactiveusers{} users of the service, 41 used the feature: 15 selected only one of the first two suggestions displayed by default (Figure~\ref{fig:navigation_suggest_initial}), 16 browsed the full list before selecting, and 10 never selected a suggestion.

When focusing on individual requests, 64\% of \ui{Suggest Follow-ups} requests resulted in the user selecting at least one suggestion. The majority of those selections (89\%) came from the first two suggestions. Follow-up suggestions were not offered on every message (for example, the \ui{Menu} message). When they were available, users chose them as their next action only 4.3\% of the time.
\section{Discussion and Lessons Learned} \label{sec:discussion}
\subsection{Design Pressures Shaping Engagement on WhatsApp}
Our design experience highlighted the distinctive pressures that WhatsApp's affordances place on engagement. WhatsApp is a closed ecosystem with a restrictive API, and its interface offers few persistent elements: messages quickly disappear in the chat scroll, and conversations lose visibility as newer ones displace them. Unlike specialized applications with richer UI components, where features remain in view, typical engagement mechanisms on WhatsApp depend on users' initiative to find them. Sustaining feature visibility in~\chatbot{} therefore required deliberate effort, such as embedding a persistent menu button in each message and using push messages to resurface the service.

Proactive communication coincided with higher short-term activity: engagement rose on broadcast days and peaked soon after TopQ messages, but most users' final sessions fell within the broadcast period. This pattern aligns with digital health studies in which reminders increased participation around each contact without sustaining it~\cite{ke2026invisible}, and with declining engagement in a prior WhatsApp deployment~\cite{whatfutures}. Because TopQ ran early in the deployment, fading novelty may also have contributed, which our data cannot separate from the end of the broadcasts. Together with prior work~\cite{ke2026invisible}, these results suggest that proactive messaging may be more sustainable when it is low-burden and aligned with users' routines.

Adoption of other engagement features was uneven across and within user groups. One-time users, the least committed by definition, used curated lists most, suggesting that peer-sourced content can offer an accessible way in for newcomers. This contrasts with prior findings that social features added to interventions mainly benefit users who are already active~\cite{ke2026invisible}, although browsing curated questions requires less effort than the social interaction those features demanded. The \ui{Leaderboard} followed the more familiar pattern: its five most frequent users were also among the service's most active, while~\servicerare{} users varied widely in their \ui{Leaderboard} activity. Together, these patterns point to a need for engagement strategies versatile enough to accommodate diverse motivations and to adapt as users' engagement evolves, particularly for general-purpose services, which lack the task-specific relevance that sustains engagement in specialized ones.

Follow-up suggestions illustrate the same trade-off. Because each message carries only three buttons, we kept the \ui{Menu} button persistent and placed suggestions behind a separate request, a choice suited to a general-purpose service whose many features require constant access to the menu. Viewing suggestions therefore required sending a request and waiting for a reply before selecting one. In more specialized chatbots, suggestions took less effort to reach: ASHABot's related questions opened directly from each response~\cite{ashabot}, and in a chatbot study, suggestion buttons appeared alongside every message and could be tapped immediately~\cite{replysuggestions}. Uptake in those systems was substantially higher, whereas~\chatbot{}'s suggestions were chosen as the next action only 4.3\% of the time. Despite these measures not being directly comparable, the pattern indicates that each additional step reduced use. Once suggestions were shown, users mostly selected the two displayed by default, reinforcing the role of visibility; suggestion quality, which we did not evaluate, may also have played a role~\cite{rosset2020leading}.

Beyond interface constraints, platform-level decisions add operational pressure. WhatsApp's restriction of third-party general-purpose chatbots shows that services built on proprietary platforms remain vulnerable to decisions beyond their control, with direct consequences for their sustainability. Although these lessons emerge from WhatsApp, several of the affordances involved, such as the prioritization of recent conversations and limited persistent interface elements, are shared by other messaging platforms, suggesting that aspects of these lessons may transfer beyond this context.

\subsection{User Perception and Interface Design}
\chatbot{}'s combination of a text-based modality and interactive buttons led to some interface friction. We observed a couple of  instances in which users appeared to conflate text-based interaction with button functionality (e.g., typing 'menu' rather than tapping the~\ui{Menu} button), though we did not systematically quantify this behavior. Prior studies have observed similar behavior among Google Bard users in the US and UK~\cite{bardAnalysis}, while others reported that users in India transferred expectations from phone calls to an audio chatbot~\cite{farmchat}. Users from contrasting backgrounds can thus experience similar friction when navigating the duality between UI elements and chatbot functionality.

Language expectations further shaped user experience. Users occasionally explored whether ~\chatbot{} could understand their native languages. LLMs still struggle with non-English text generation~\cite{changLlmssurvey}. Combined with the translation requests observed in our dataset, this limitation underscores the importance of linguistic accessibility as a key enabler of broader adoption.

Together, these observations suggest that trust and adoption depend not only on accuracy but also on alignment with users' linguistic needs, cultural expectations, and mental models~\cite{ai-cultural}. Supporting both UI-based and natural language interaction can improve accessibility and intuitiveness.

\subsection{Broad Use and Contextual Settings} 
Our topic distribution overlaps with other work about the use of ChatGPT in the wild~\cite{peopleusegpt} in common categories such as learning, translation, and communication. However, health and well-being, our largest category, is far less prominent in their global dataset (only 6.7\%). These differences could be influenced by the context within which WhatsApp and mobile devices are used.

Users turned to~\chatbot{} for factual information, verification, and critical topics such as health and nutrition, suggesting that it may serve as an accessible entry point where other resources are less convenient. Although inaccurate responses were uncommon and mostly concerned everyday topics, most went unchallenged, raising concerns about uncritical acceptance of plausible-sounding errors~\cite{ji2023survey,changLlmssurvey}, particularly as users also brought high-stakes questions to the service. Prior work suggests that interface features such as reply suggestions can increase users' willingness to follow chatbot advice~\cite{replysuggestions}. Making such features more visible may therefore support engagement while also encouraging acceptance, suggesting that they should be paired with cues that prompt verification of critical information.

\subsection*{Limitations and Future Work}
\chatbot{} served as a design probe rather than an evaluation of specific design choices, highlighting challenges in adapting engagement mechanisms to constrained contexts such as WhatsApp. Our findings are limited by an English-only interface, a small user base, limited demographic data, and a focus on early-phase use. Future work should examine predictors of sustained engagement and evaluate specific design choices, such as the quality of suggested questions.

\begin{acks}
Thanks to the Tufts NAT lab and D.O.C.C. lab as well as all users of our WhatsApp Q\&A service for their feedback and support. \\This work was partially funded by NSF CNS award: 2106797.
\end{acks}
\bibliographystyle{ACM-Reference-Format}
\bibliography{main}
%
\appendix
\section{Prompts}
\label{appendix-prompts}
Each subsection title below describes a specific task performed by the LLM and contains a pair of System Prompt and User Prompt,
which are together used as input (prompt) to the LLM. The prompts guide the LLM to perform specific tasks.
\subsection{Prompt for answering a user's query.}
\textbf{\\\hspace*{1em}System Prompt:}

\begin{tcolorbox}[breakable, colback=gray!5, colframe=gray!40, boxrule=0.5pt]
\small
\noindent{\ttfamily
Instructions

You are an advanced AI assistant designed to provide informative and helpful responses to a wide range of queries.

Your responses should be clear and concise, adhering to the following guidelines:

1. You can use the context provided but only to get the information needed to address the user's current message.\\
2. No need to follow the length of responses, verbosity etc. of the messages in the context.\\
3. Engage in a conversational manner and use humor if applicable.\\
4. Your responses should be visually appealing for WhatsApp users, so use emojis, short paragraphs, etc.
}
\end{tcolorbox}

\textbf{User Prompt:}

\begin{tcolorbox}[breakable, colback=gray!5, colframe=gray!40, boxrule=0.5pt]
\small\ttfamily
\{user-query\}
\end{tcolorbox}

\subsection{Prompt for generating a more detailed answer to a user's query using a high-quality model.}
\textbf{\\\hspace*{1em}System Prompt:}

\begin{tcolorbox}[breakable, colback=gray!5, colframe=gray!40, boxrule=0.5pt]
\small
\noindent{\ttfamily
Instructions

You are an advanced AI assistant designed to provide informative and helpful responses to a wide range of queries.

Your response should be detailed and thoughtful and go beyond simply answering the user's query.

1. You can use the context provided but only to get the information needed to address the user's current message.\\
2. No need to follow the length of responses, verbosity etc. of the messages in the context.\\
3. Engage in a conversational manner and use humor if applicable.\\
4. Your responses should be visually appealing for WhatsApp users, so use emojis, short paragraphs, etc.
}
\end{tcolorbox}

\textbf{User Prompt:}

\begin{tcolorbox}[breakable, colback=gray!5, colframe=gray!40, boxrule=0.5pt]
\small\ttfamily
\{user-query\}
\end{tcolorbox}

\subsection{Prompt for generating follow-up questions based on a user's query.}

\textbf{\\\hspace*{1em}System Prompt:}

\begin{tcolorbox}[breakable, colback=gray!5, colframe=gray!40, boxrule=0.5pt]
\small
\noindent{\ttfamily
Instructions

You are given a user's query and the original response from a chatbot.

Your task is to formulate 6 follow-up questions that THE USER CAN ASK further based on things the response hasn't covered. These questions should pique the user's interest, with the goal to make them want to keep using the chatbot.

Question 1 and 2 must dig deep into the topic of the given query, focusing on some specific detail or aspect of it.\\
Question 3 and 4 must focus on a different but closely related topic to the topic of the given query.\\
Question 5 and 6 must be on a completely new topic that is loosely tied with the topic of the given query.

Format Instructions

You must adhere to the following:
1. Each question should be concise, ideally 1 sentence long.\\
2. Each question statement should be visually appealing for WhatsApp users, so you must use emojis in the question statement.\\
3. Strictly respond in the following JSON format: \{"q1": "question statement", "q2": "question statement", "q3": "question statement", "q4": "question statement", "q5": "question statement", "q6": "question statement"\}
}
\end{tcolorbox}

\textbf{User Prompt:}

\begin{tcolorbox}[breakable, colback=gray!5, colframe=gray!40, boxrule=0.5pt]
\small\ttfamily
\{user-query\}
\end{tcolorbox}

\subsection{Prompt for deciding whether or not a question should be added to Recent Questions' list.}
\textbf{\\\hspace*{1em}System Prompt:}

\begin{tcolorbox}[breakable, colback=gray!5, colframe=gray!40, boxrule=0.5pt]
\small
\noindent{\ttfamily
Instructions

You are an advanced AI designed to help the user improve and analyse language usage.
}
\end{tcolorbox}

\textbf{User Prompt:}

\begin{tcolorbox}[breakable, colback=gray!5, colframe=gray!40, boxrule=0.5pt]
\small\ttfamily
Statement: \{user-query\}

Above, you are provided with a statement by a user. DO NOT ATTEMPT TO ANSWER IT. Perform the following analysis on it:

1. Return None if the statement is not a question statement.\\
2. Return None if the question statement is in any language other than English.\\
3. Return None if the question statement refers to any object/idea/thing/text etc. that is not explicitly defined within the question statement.\\
4. Convert it to a SINGLE question statement.\\
5. Fix any typos in it and make it less than 125 words.\\
6. Add a relevant emoji to it to make it visually appealing.

ONLY return None or the rephrased question statement.
\end{tcolorbox}

\subsection{Prompt for rating a question to decide whether or not to include it in Trending Questions list.}
\textbf{\\\hspace*{1em}System Prompt:}

\begin{tcolorbox}[breakable, colback=gray!5, colframe=gray!40, boxrule=0.5pt]
\small\ttfamily
Instructions

You are an advanced AI data analyst designed to help the user analyze user questions.
\end{tcolorbox}

\textbf{User Prompt:}

\begin{tcolorbox}[breakable, colback=gray!5, colframe=gray!40, boxrule=0.5pt]
\small\ttfamily
\{user-query\}\\
I am providing a criteria to you to rate questions. Assign a binary score (0/1) for the provided question for each criteria point. 1 indicates the question meets the criteria point, while 0 indicates that it does not. There are a total of 10 criteria points. The question is: \{question-statement\}. The criteria is:

\begin{itemize}[leftmargin=1cm, label={}]
    \item criteria\_1 [In English]: Question is written in the English language.
    \item criteria\_2 [Independent]: The question does not refer to any object/idea/thing/text etc. that is not explicitly defined within the query statement.
    \item criteria\_3 [General Interest Topics]: Question is about a widely recognized subject such as common products, health, finance, technology, education, entertainment, or current events.
    \item criteria\_4 [Recurring Themes]: Question touches on common human experiences or needs, such as relationship or career advice, general guidance, parenting, hobbies or personal growth.
    \item criteria\_5 [Intellectually Stimulating]: Question is intellectually stimulating.
    \item criteria\_6 [Funny But Interesting]: Question is funny but poses an interesting question at the same time.
    \item criteria\_7 [Curiosity]: Question is interesting and will likely pique one's curiosity.
    \item criteria\_8 [Life Stages and Milestones]: Question is about life events that most people go through, such as schooling, starting a job, marriage, retirement, etc.
    \item criteria\_9 [Problem-Solving Orientation]: Question is around solving a common problem or dealing with a situation that many might face, regardless of demographic.
    \item criteria\_10 [Broadly Applicable Advice]: Question is on asking for advice or recommendations which could be useful for a wide range of people, not limited by geography, age, or occupation.
\end{itemize}

Your response should only contain comma-separated scores for each criteria point 1-10 in the same order: score1,score2,score3,score4,...,score10
\end{tcolorbox}

\subsection{Prompt for preprocessing a question before it is added to Trending or Recent Questions lists.}
\textbf{\\\hspace*{1em}System Prompt:}

\begin{tcolorbox}[breakable, colback=gray!5, colframe=gray!40, boxrule=0.5pt]
\small\ttfamily
Instructions\\
You are an advanced AI designed to help the user rephrase question statements.
\end{tcolorbox}

\textbf{User Prompt:}

\begin{tcolorbox}[breakable, colback=gray!5, colframe=gray!40, boxrule=0.5pt]
\small\ttfamily
\{question-statement\}

You are provided with a question statement. Rephrase the question following this criteria:

1. Fix any typos or punctuation errors.\\
2. Add one emoji to the question statement to make it visually appealing.\\
3. Do \textbf{NOT} answer the question and do \textbf{NOT} add unnecessary details to the question statement.\\
4. ONLY if the question is longer than 150 words, shorten it.

\noindent Strictly return \textbf{ONLY} the rephrased question.
\end{tcolorbox}

\subsection{Prompt for generating answers to follow-up, trending and recent questions.}
\textbf{\\\hspace*{1em}System Prompt:}

\begin{tcolorbox}[breakable, colback=gray!5, colframe=gray!40, boxrule=0.5pt]
\small
\noindent{\ttfamily
Instructions

You are an advanced AI assistant designed to provide informative and helpful responses to a wide range of queries.

Your responses should be clear and concise.

Format Instructions

1. If the query can be answered concisely, do not be verbose. For example, factual queries can be answered in a few sentences.\\
2. Engage in a conversational manner and use humor if applicable.\\
3. Your responses should be visually appealing for WhatsApp users, so use emojis, short paragraphs, etc.
}
\end{tcolorbox}

\textbf{User Prompt:}

\begin{tcolorbox}[breakable, colback=gray!5, colframe=gray!40, boxrule=0.5pt]
\small\ttfamily
\{user-query\}
\end{tcolorbox}
\section{Top-Rated Questions per User Group}
\label{appendix-trending-questions}
\begin{table}[H]
\caption{\servicefrequentcaps{} users top-rated questions}
\label{tab:top_trending_queries_p}
\begin{tabular} {
    |>{\raggedright\arraybackslash}m{0.45\linewidth}
    |>{\centering\arraybackslash}m{0.25\linewidth}
    |>{\centering\arraybackslash}m{0.12\linewidth}
    |
}
\hline
\multicolumn{3}{|c|}{\textbf{\servicefrequentcaps{} Users}}\\
\hline
\makecell[c]{Top-rated Questions} & \makecell[c]{Topic\\Category} & \makecell[c]{Chosen\\x Times}\\
\hline
How can we provide emotional care for our parents in their old age? & Social and Personal Development & 7\\ \hline
In which fields and subjects is a PhD worth pursuing in this day and age?  & Education and Learning & 5 \\ \hline
I used to be able to memorize the names of every student in my class, but that is no longer the case. What can I do? & Social and Personal Development & 4 \\ \hline
What are the main reasons for divorce around the world? & Social and Personal Development & 4 \\ \hline
Is becoming a doctor suitable for everyone, considering the finances, intellect, and time required?  & Social and Personal Development & 4 \\ \hline
How do marriage and children affect a woman's career? & Social and Personal Development & 4\\ \hline
\end{tabular}
\end{table}

\begin{table}[H]
\caption{\servicerarecaps{} users top-rated questions}
\label{tab:top_trending_queries_i}
\begin{tabular} {
    |>{\raggedright\arraybackslash}m{0.45\linewidth}
    |>{\centering\arraybackslash}m{0.25\linewidth}
    |>{\centering\arraybackslash}m{0.12\linewidth}
    |
}
\hline
\multicolumn{3}{|c|}{\textbf{\servicerarecaps{} Users}}\\
\hline
\makecell[c]{Top-rated Questions} & \makecell[c]{Topic\\Category} & \makecell[c]{Chosen\\x Times}\\
\hline
What are the psychological effects on children of broken marriages?  & Health and Well-being & 9\\ \hline
Why do we say "sleep like a baby" when babies wake up crying every few hours? & Language and Communication & 9\\ \hline
Are late marriages better than early marriages?  & Social and Personal Development  & 9 \\ \hline
Is it okay for parents to push their children into mastering a skill from an early age, or should children be given the freedom to pursue their own interests? & Social and Personal Development & 8 \\ \hline
How can we provide emotional care for our parents in their old age?  & Social and Personal Development & 8\\ \hline
What are the main reasons for divorce around the world? & Social and Personal Development & 8 \\ \hline
What are effective ways to teach children patience? & Social and Personal Development & 8 \\ \hline
What steps can individuals take to successfully transition to a new career later in life?  & Social and Personal Development & 8 \\ \hline
How can we provide emotional care for our parents in their old age? & Social and Personal Development & 8\\ \hline
What are some ways to better prepare for retirement? & Social and Personal Development & 8 \\ \hline
\end{tabular}
\end{table}
\end{document}